\documentclass[aps,prd,onecolumn,groupedaddress,showpacs,nofootinbib,amssymb]{revtex4}
\usepackage[dvips]{graphicx}
\usepackage{amssymb}
\usepackage{amsmath}
\usepackage{graphicx,,color}
\usepackage{amsfonts}
\usepackage{bm}
\usepackage{cancel}
\usepackage{comment}

\newcommand\be{\begin{equation}}
\newcommand\ee{\end{equation}}

\allowdisplaybreaks[4]

\begin{document}

\tolerance=5000

\title{Late-Time Cosmology of Scalar-Coupled $f(R, \mathcal{G})$ Gravity}
\author{S.D.~Odintsov,$^{1,2}$\,\thanks{odintsov@ieec.uab.es}
V.K.~Oikonomou,$^{3,4}$\,\thanks{v.k.oikonomou1979@gmail.com}F.P.
Fronimos,$^{3}$\,\thanks{fotisfronimos@gmail.com}}
\affiliation{$^{1)}$ ICREA, Passeig Luis Companys, 23, 08010 Barcelona, Spain\\
$^{2)}$ Institute of Space Sciences (IEEC-CSIC) C. Can Magrans
s/n,
08193 Barcelona, Spain\\
$^{3)}$ Department of Physics, Aristotle University of
Thessaloniki, Thessaloniki 54124,
Greece\\
$^{4)}$ Laboratory for Theoretical Cosmology, Tomsk State
University of Control Systems and Radioelectronics, 634050 Tomsk,
Russia (TUSUR)}

\tolerance=5000

\begin{abstract}
In this work by using a numerical analysis, we investigate in a
quantitative way the late-time dynamics of scalar coupled
$f(R,\mathcal{G})$ gravity. Particularly, we consider a
Gauss-Bonnet term coupled to the scalar field coupling function
$\xi(\phi)$, and we study three types of models, one  with $f(R)$
terms that are known to provide a viable late-time phenomenology,
and two Einstein-Gauss-Bonnet types of models. Our aim is to write
the Friedmann equation in terms of appropriate statefinder
quantities frequently used in the literature, and we numerically
solve it by using physically motivated initial conditions. In the
case that $f(R)$ gravity terms are present, the contribution of
the Gauss-Bonnet related terms is minor, as we actually expected.
This result is robust against changes in the initial conditions of
the scalar field, and the reason is the dominating parts of the
$f(R)$ gravity sector at late times. In the Einstein-Gauss-Bonnet
type of models, we examine two distinct scenarios, firstly by
choosing freely the scalar potential and the scalar Gauss-Bonnet
coupling $\xi(\phi)$, in which case the resulting phenomenology is
compatible with the latest Planck data and mimics the
$\Lambda$-Cold-Dark-Matter model. In the second case, since there
is no fundamental particle physics reason for the graviton to
change its mass, we assume that primordially the tensor
perturbations propagate with the speed equal to that of light's,
and thus this constraint restricts the functional form of the
scalar coupling function $\xi(\phi)$, which must satisfy the
differential equation $\ddot{\xi}=H\dot{\xi}$. The latter equation
is greatly simplified when late times are considered and can be
integrated analytically to yield a relation for $\dot{\xi}$, which
depends solely on the Hubble rate, in a model independent way.
This leads eventually to an elegant simplification of the
Friedmann equation, which when solved numerically, yields a viable
late-time phenomenology. A common characteristic of the
Einstein-Gauss-Bonnet models we considered is that the dark energy
era they produce is free from dark energy oscillations.
\end{abstract}


\maketitle

\section{Introduction}

The quest for understanding the mysterious late-time acceleration
era \cite{Riess:1998cb}, is still ongoing in modern theoretical
physics. Many possible theoretical descriptions have been proposed
in order to model the dark energy era, among which modified
gravity has an elevated role in the successful description of the
dark energy era
\cite{Nojiri:2017ncd,Nojiri:2009kx,Capozziello:2011et,Capozziello:2010zz,Nojiri:2006ri,
Nojiri:2010wj,delaCruzDombriz:2012xy,Olmo:2011uz}, since apart
from being able to describe the late-time era, it is also possible
to describe inflation with the same theoretical framework, see for
example Refs.
\cite{Nojiri:2003ft,Nojiri:2007as,Nojiri:2007cq,Cognola:2007zu,Nojiri:2006gh,Appleby:2007vb,Elizalde:2010ts,Odintsov:2020nwm}.
Modern modified gravity models are put into stringent test of
viability when the dark energy era is considered, since the models
must be confronted with the latest Planck 2018 data
\cite{Aghanim:2018eyx}, and also the models have to be compatible
to some inferior extent with the $\Lambda$-Cold-Dark-Matter
($\Lambda$CDM) model, which is the most successful model for
describing the dark energy era. The $\Lambda$CDM model is
basically based on the assumptions of the presence of a
cosmological constant, and the presence of particle dark matter
\cite{Bertone:2004pz,Bergstrom:2000pn,Mambrini:2015sia,Profumo:2013yn,Hooper:2007qk,Oikonomou:2006mh},
however both the ingredients of the model are in question of their
existence. Moreover, although the $\Lambda$CDM is quite compatible
with the Cosmic Microwave Background data, it has several
theoretical shortcomings which cannot be harbored by
Einstein-Hilbert gravity. At this point, modified gravity offers
many possibilities for successful theoretical descriptions. In
this line of research, Einstein-Gauss-Bonnet theory
\cite{Hwang:2005hb,Nojiri:2006je,Cognola:2006sp,Nojiri:2005vv,Nojiri:2005jg,Satoh:2007gn,Bamba:2014zoa,Yi:2018gse,Guo:2009uk,Guo:2010jr,Jiang:2013gza,Kanti:2015pda,vandeBruck:2017voa,Kanti:1998jd,Pozdeeva:2020apf,Fomin:2020hfh,DeLaurentis:2015fea,Chervon:2019sey,Nozari:2017rta,Odintsov:2018zhw,Kawai:1998ab,Yi:2018dhl,vandeBruck:2016xvt,Kleihaus:2019rbg,Bakopoulos:2019tvc,Maeda:2011zn,Bakopoulos:2020dfg,Ai:2020peo,Odintsov:2019clh,Oikonomou:2020oil,Odintsov:2020xji,Oikonomou:2020sij,Odintsov:2020zkl,Odintsov:2020sqy,Odintsov:2020mkz,Easther:1996yd,Antoniadis:1993jc,Antoniadis:1990uu,Kanti:1995vq,Kanti:1997br}
could be a potentially correct description of both the early and
late-time era. In this work we shall investigate in a quantitative
way the exact effect of the Gauss-Bonnet coupling on the late-time
era, in the context of $f(R,\phi)$ theories of gravity in general.
In particular, we shall investigate the effect of the non-trivial
Gauss-Bonnet coupling on a simple canonical scalar field theory,
and on $f(R)$ gravity in the presence of a canonical scalar field.
We shall perform a thorough numerical analysis of the Friedmann
equation and we shall derive the behavior of several statefinder
quantities and of several physical quantities of interest, as
functions of the redshift. Accordingly our findings shall be
compared with the $\Lambda$CDM and shall be confronted with the
latest Planck constraints on the cosmological parameters
\cite{Aghanim:2018eyx}. Our findings indicate that when an $f(R)$
gravity theory is present along with the Gauss-Bonnet coupling,
the latter does not significantly affect the late-time
phenomenology. Also in the case of a simple Einstein-Gauss-Bonnet
theory, we show that it is possible to obtain the phenomenological
viability of the model under study, but this result could be
highly model dependent, and has a minor disadvantage, since it is
hard to describe inflation and dark energy with the same
Einstein-Gauss-Bonnet model, in general though.

Finally, we make a novel assumption that may constrain the
functional form of the scalar coupling function of the scalar
field to the Gauss-Bonnet coupling, and we investigate the
late-time phenomenology in this case too. Particularly, we assume
that there is a constraint coming from the requirement that the
primordial gravitational wave speed is equal to unity, which
imposes a functional constraint on the functional form of the
Gauss-Bonnet scalar coupling function. The reason for demanding
that the primordial gravitational wave speed is equal to unity in
natural units, is coming from the GW170817 event
\cite{GBM:2017lvd}, which indicated that the gamma rays and the
gravitational waves arrived almost simultaneously. Thus assuming
that there is no fundamental particle physics reason for the
graviton to change its primordial mass, the gravity speed
constraint imposed by the GW170817 event, must stretch back to the
primordial inflationary era. For the late-time era, the constraint
imposed by requiring that the gravity wave speed of the primordial
tensor modes is equal to unity in natural units, results to an
elegant expression for the time derivative of the scalar
Gauss-Bonnet coupling, which is expressed in terms of the Hubble
rate and the redshift, and thus it depends on statefinder
quantities and acquires a model independent description. The
late-time viability of such an Einstein-Gauss-Bonnet gravity is
examined in detail, and our findings indicate that these can also
provide a successful description of the late-time era. Finally, we
also conclude that the Einstein-Gauss-Bonnet theories in general,
provide a dark energy oscillations free late-time era, in contrast
to $f(R)$ gravity models. However, this result seems to be highly
model dependent, at least in the context of $f(R)$ gravity and
needs to be further discussed in another context.

\section{Essential Features of $f(R,\phi)$ Einstein-Gauss-Bonnet Gravity}

The starting point of our work is the gravitational action, which
for the $f(R,\phi)$ Einstein-Gauss-Bonnet gravity has the
following form,
\begin{equation}
\centering
\label{action}
S=\int{d^4x\sqrt{-g}\left(\frac{f(R,\phi)}{2\kappa^2}-\frac{1}{2}g^{\mu\nu}\partial_\mu\phi\partial_\nu\phi-V-\xi(\phi)\mathcal{G}+\mathcal{L}_{(m)}\right)}\, ,
\end{equation}
where $R$ denotes the Ricci scalar, $\kappa=\frac{1}{M_P}$ is the
gravitational constant with $M_P$ being the reduced Planck mass,
while $V$ is the scalar potential, while $\xi(\phi)$ is the
Gauss-Bonnet scalar coupling function. Also
$\mathcal{G}=R^2-4R_{\mu\nu}R^{\mu\nu}+R_{\mu\nu\sigma\rho}R^{\mu\nu\sigma\rho}$
denotes the Gauss-Bonnet invariant with $R_{\mu\nu}$ and
$R_{\mu\nu\sigma\rho}$ being the Ricci and Riemann curvature
tensor respectively and finally, $\mathcal{L}_{(m)}$ specifies the
Lagrangian density of both relativistic and non-relativistic
perfect matter fluids. For now, the exact form of the function
$f(R,\phi)$ shall remain unspecified, at least for the moment.
Concerning the cosmological geometric background, we shall assume
that it corresponds to that of a flat Friedman-Robertson-Walker
(FRW) metric with the line element being,
\begin{equation}
\centering
\label{metric}
ds^2=-dt^2+a(t)^2\delta_{ij}dx^idx^j\, ,
\end{equation}
where $a(t)$ denotes the scale factor. Consequently, the Ricci
scalar and Gauss-Bonnet invariant for the FRW background take the
forms $R=6(2H^2+\dot H)$ and $\mathcal{G}=24H^2(H^2+\dot H)$,
where $H=\frac{\dot a}{a}$ signifies the Hubble rate and the
``dot'' denotes differentiation with respect to the cosmic time
$t$. Furthermore, in order to simplify our work, we shall make the
reasonable assumption that the scalar field is homogeneous and
thus it depends solely on the cosmic time.

Implementing the variation principle with respect to the metric
tensor $g^{\mu\nu}$ and the scalar field $\phi$, we obtain the
field equations for gravitational sector and the scalar field
equation, which are,
\begin{equation} \centering \label{motion1}
\frac{3f_RH^2}{\kappa^2}=\rho_{(m)}+\frac{1}{2}\dot\phi^2+V+\frac{f_R
R-f}{2\kappa^2}-\frac{3H\dot f_R}{\kappa^2}+24\dot\xi H^3\, ,
\end{equation}
\begin{equation}
\centering
\label{motion2}
-\frac{2f_R\dot H}{\kappa^2}=\rho_{(m)}+P_{(m)}+\dot\phi^2+\frac{\ddot f_R-H\dot f_R}{\kappa^2}-16\dot\xi H\dot H\, ,
\end{equation}
\begin{equation}
\centering
\label{motion3}
V_\phi+\ddot\phi+3H\dot\phi-\frac{f_\phi}{2\kappa^2}+\xi_{\phi}\mathcal{G}=0\, ,
\end{equation}
where $\rho_{(m)}$ and $P_{(m)}$ denote the matter density and
pressure respectively of any prefect fluid of non-relativistic
(baryons, leptons, Cold Dark matter (CDM)) matter and relativistic
matter (photons and neutrinos). In particular, for the purposes of
this work, for which the focus will be on the dark energy era, we
shall assume that the perfect fluids compose of CDM and radiation,
so we have,
\begin{equation}
\centering
\label{rho}
\rho_{(m)}=\rho_{d0}\frac{1}{a^3}\left(1+\chi\frac{1}{a}\right)\, ,
\end{equation}
\begin{equation}
\centering
\label{P}
P=\sum_{i}{\omega_i\rho_i}\, ,
\end{equation}
where $\chi=\frac{\rho_{r0}}{\rho_{d0}}$ with $\rho_{r0}$ being
the current density of relativistic matter and $\omega_i$ the
equation of state parameter for each kind or matter, with $i$
running from relativistic to non-relativistic. As we already
mentioned, all matter species are described by perfect fluids and
a barotropic EoS. Since we have prefect fluids, the continuity
equation for each of them reads,
\begin{equation}
\centering \label{conteq} \dot\rho_i+3H\rho_i(1+\omega_i)=0\, .
\end{equation}
In the following, we shall implement two replacements in order to
align better our study with the late-time dynamics, with the first
being the use of the redshift instead of the cosmic time as a
dynamical parameter. From its definition,
\begin{equation}
\centering
\label{z}
1+z=\frac{1}{a(t)}\, ,
\end{equation}
where we assumed that the scale factor at present time is set to
unity, the time derivatives can be expressed in terms of
derivatives with respect to the redshift, using the following
rule,
\begin{equation}
\centering
\label{dz}
\frac{d}{dt}=-H(1+z)\frac{d}{dz}\, ,
\end{equation}
By replacing cosmic time $t$ with the redshift, one can recast the
cosmological equations of motion in the following
way,
\begin{equation} \centering \label{motion4}
\frac{3f_RH^2}{\kappa^2}=\rho_{(m)}+\frac{1}{2}\dot\phi^2+V+\frac{f_R
R-f}{2\kappa^2}-\frac{3H\dot f_R}{\kappa^2}-24(1+z)\xi' H^4\, ,
\end{equation}
\begin{equation}
\centering
\label{motion5}
V_\phi+\ddot\phi+3H\dot\phi-\frac{f_\phi}{2\kappa^2}+\xi_\phi\mathcal{G}=0\, ,
\end{equation}
where the ``prime'' denotes differentiation with respect to the
redshift. Here, only two equations where rewritten since they will
be used in our study. In addition, every time derivative
participating in the equations of motion above shall be replaced
as well. Specifically, we have,
\begin{equation}
\centering
\dot H=-H(1+z)H'\, ,
\end{equation}
\begin{equation}
\centering
\dot\phi=-H(1+z)\phi'\, ,
\end{equation}
\begin{equation}
\centering
\ddot\phi=H^2(1+z)^2\phi''+H^2(1+z)\phi'+HH'(1+z)^2\phi'\, ,
\end{equation}
\begin{equation}
\centering
\dot f_R=\dot Rf_{RR}+\dot\phi f_{R\phi}\, ,
\end{equation}
\begin{equation}
\centering
\dot R=6H(1+z)^2\left(HH''+(H')^2-\frac{3HH'}{1+z}\right)\, ,
\end{equation}
Now more importantly, instead of using the Hubble rate and its
derivatives in order to quantify the cosmological evolution, we
shall use a statefinder quantity defined as follows
\cite{Hu:2007nk,Bamba:2012qi,Odintsov:2020qyw,Odintsov:2020nwm},
\begin{equation}
\centering
\label{yH}
y_H=\frac{\rho_{DE}}{\rho_{d0}}\, ,
\end{equation}
with $\rho_{DE}$ denoting the dark matter energy density and
$\rho_{d0}$ the current value of density for non-relativistic
matter. Here, we shall assume that the dark energy density is
comprised of all the geometric terms in the Friedmann equation. In
particular,
\begin{equation}
\centering
\label{DEdensity}
\rho_{DE}=\frac{1}{2}\dot\phi^2+V+\frac{f_R R-f}{2\kappa^2}-\frac{3H\dot f_R}{\kappa^2}+24\dot\xi H^3+\frac{3H^2}{\kappa^2}(1-f_R)\, ,
\end{equation}
Similarly, from the Raychaudhuri equation, the corresponding
pressure for the dark energy fluid is defined as,
\begin{equation}
\centering
\label{PDE}
P_{DE}=-V-24\dot\xi H^3-8\dot\xi H\dot H-\frac{f_R R-f}{2\kappa^2}-\frac{2\dot H}{\kappa^2}(1-f_R)\, ,
\end{equation}
where,
\begin{equation}
\centering
\label{conteqDE}
\dot\rho_{DE}+3H(\rho_{DE}+P_{DE})=0\, ,
\end{equation}
Hence, equations (\ref{motion1}) and (\ref{motion2}) obtain the
usual Friedmann equation-like form of Einstein-Hilbert gravity,
\begin{equation}
\centering
\label{motion6}
\frac{3H^2}{\kappa^2}=\rho_{(m)}+\rho_{DE}\, ,
\end{equation}
\begin{equation}
\centering
\label{motion7}
-\frac{2\dot H}{\kappa^2}=\rho_{(m)}+P_{(m)}+\rho_{DE}+P_{DE}\, ,
\end{equation}
Consequently, the newly defined statefinder parameter $y_H$ can be
written in terms of the Hubble rate, and vice-versa. Specifically,
we have,
\begin{equation}
\centering
\label{H}
H^2=m_s^2\left(y_H(z)+\frac{\rho_{(m)}}{\rho_{d0}}\right)\, ,
\end{equation}
where $m_s^2=\kappa^2\frac{\rho_{d0}}{3}=1.87101\cdot10^{-67}$.
This extends to the derivatives of the Hubble rate as well since
now,
\begin{equation}
\centering
\label{H'}
HH'=\frac{m_s^2}{2}\left(y_H'+\frac{\rho_{(m)}'}{\rho_{d0}}\right)\, ,
\end{equation}
\begin{equation}
\centering
\label{H''}
H'^2+HH''=\frac{m_s^2}{2}\left(y_H''+\frac{\rho_{(m)}''}{\rho_{d0}}\right)\, ,
\end{equation}
In the following, we shall numerically solve the system of
differential equations (\ref{motion4}) and (\ref{motion5}) with
respect to the statefinder quantity $y_H$ and the scalar field
$\phi$. Afterwards, we shall compare the theoretical results with
the observations. This can be achieved by utilizing further
statefinder parameters. Concerning dark energy, we define the
equation of state parameter $\omega_{DE}$ and the density
parameter $\Omega_{DE}$ with respect to z and $y_H$ as follows
\cite{Bamba:2012qi,Odintsov:2020qyw,Odintsov:2020nwm},
\begin{align}
\centering
\label{DE}
\omega_{DE}&=-1+\frac{1+z}{3}\frac{d\ln{y_H}}{dz}&\Omega_{DE}&=\frac{y_H}{y_H+\frac{\rho_{(m)}}{\rho_{d0}}}\, ,
\end{align}
Furthermore, for the overall evolution, we shall use the following
statefinder parameters \cite{Odintsov:2020qyw,Odintsov:2020nwm},
\begin{align}
\centering
q&=-1-\frac{\dot H}{H^2}&j&=\frac{\ddot H}{H^2}-3q-2&s&=\frac{j-1}{3\left(q-\frac{1}{2}\right)}&Om(z)&=\frac{\left(\frac{H}{H_0}\right)^2-1}{(1+z)^3-1}\, ,
\end{align}
which in the order of appearance above, are the deceleration
parameter, the jerk, the snap parameter and $Om(z)$, which is
indicative of the current CDM energy density parameter.

\section{$f(R)$ Einstein-Gauss-Bonnet Gravity: Unifying Early And Late Time}

Let us commence our study by introducing the arbitrary functions
of the previous models. Hereafter, we shall limit our work to only
simple cases for the scalar functions, namely $V(\phi)$ and
$\xi(\phi)$, for which it is known that the early-time can be
described successfully. Suppose that the scalar functions of the
previous section obtain the following forms, which are arbitrary
for the moment, meaning that there is no fundamental relation
between the scalar potential and the scalar coupling function,
\begin{figure}[h!]
\centering
\label{plot1}
\includegraphics[width=20pc]{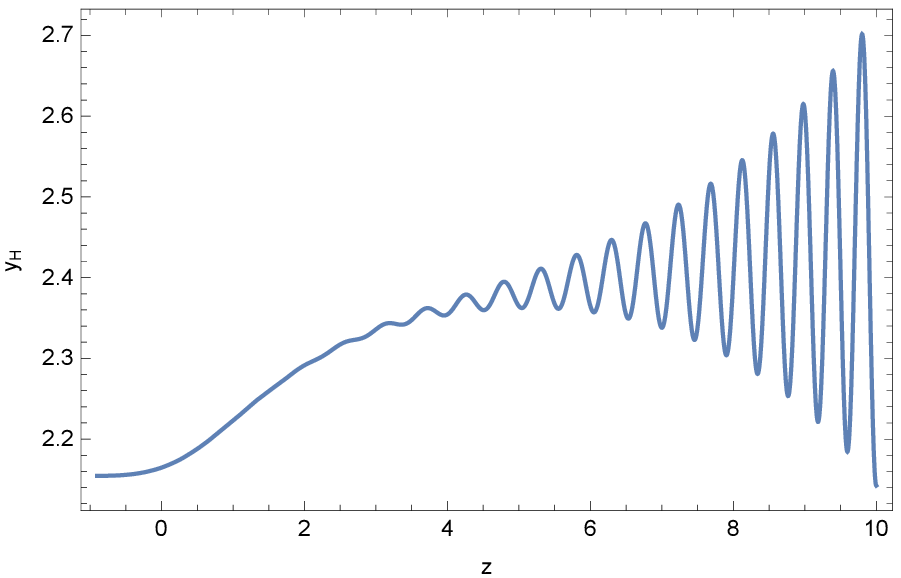}
\includegraphics[width=20pc]{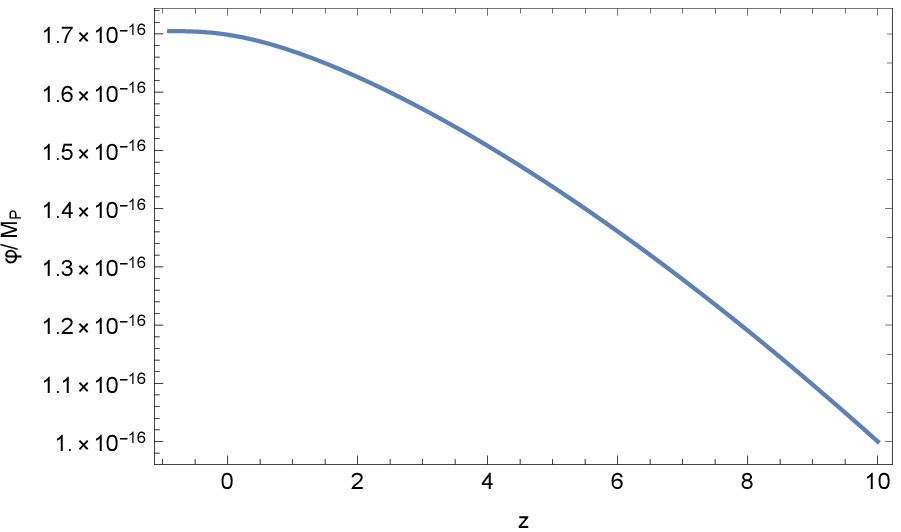}
\caption{Solutions $y_H$ (left) and $\phi$ over reduced Planck
mass (right) for the $f(R)$ case. The main difference seems to be
the scalar field which does not oscillate. In general, the
addition of a canonical scalar field and a linear Gauss-Bonnet
topological invariant coupled to a scalar function do not suffice
to nullify dark energy oscillation at large redshifts.}
\end{figure}
\begin{equation}
\centering
\label{xiA}
\xi(\phi)=e^{\frac{\phi}{M_P}}\, ,
\end{equation}
and also assuming that there is no scalar potential present, we
assume that the there is also an $f(R)$ gravity part present too,
and has the form \cite{Odintsov:2020qyw,Odintsov:2020nwm},
\begin{equation}
\centering
\label{fR1}
f(R)=R+\left(\frac{R}{M}\right)^2-\gamma\Lambda\left(\frac{R}{3m_s^2}\right)^\delta\, ,
\end{equation}
\begin{figure}[h!]
\centering
\label{plot2}
\includegraphics[width=20pc]{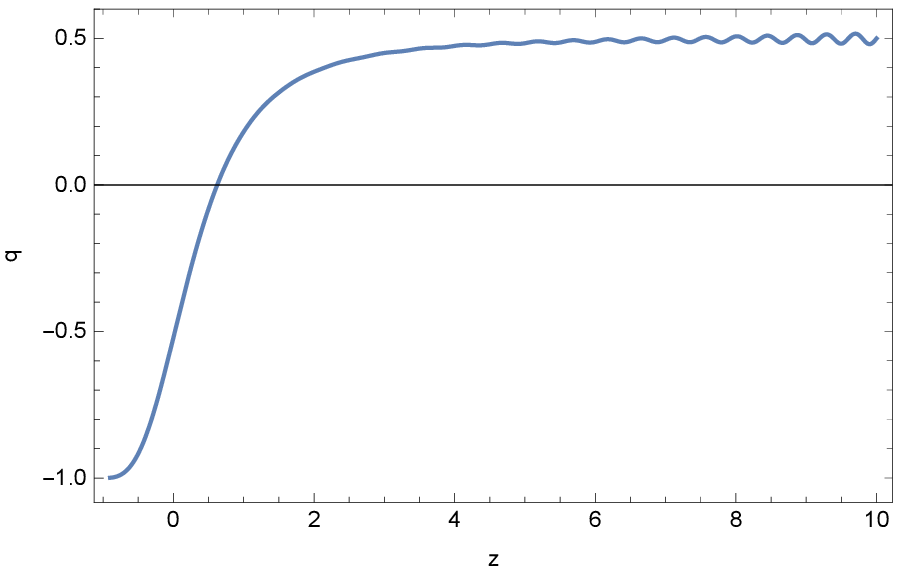}
\includegraphics[width=20pc]{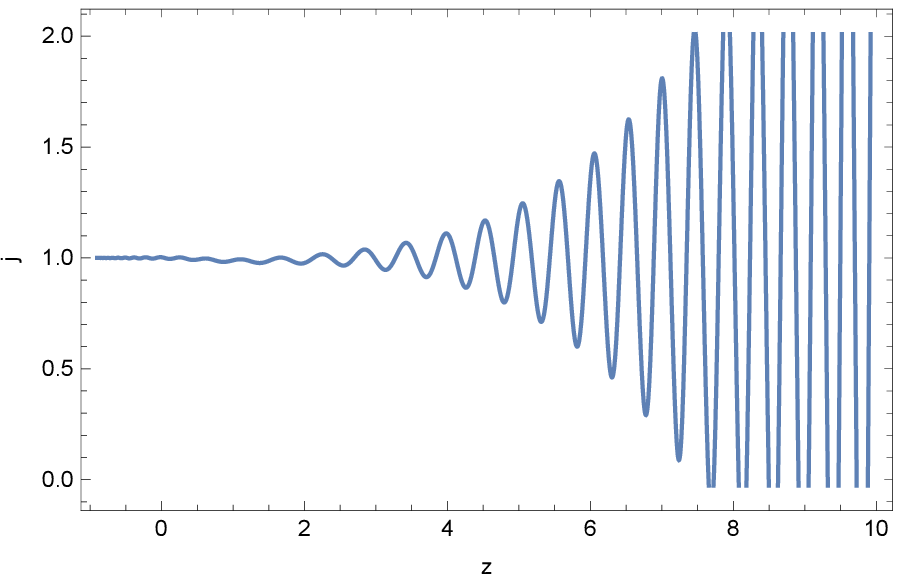}
\includegraphics[width=20pc]{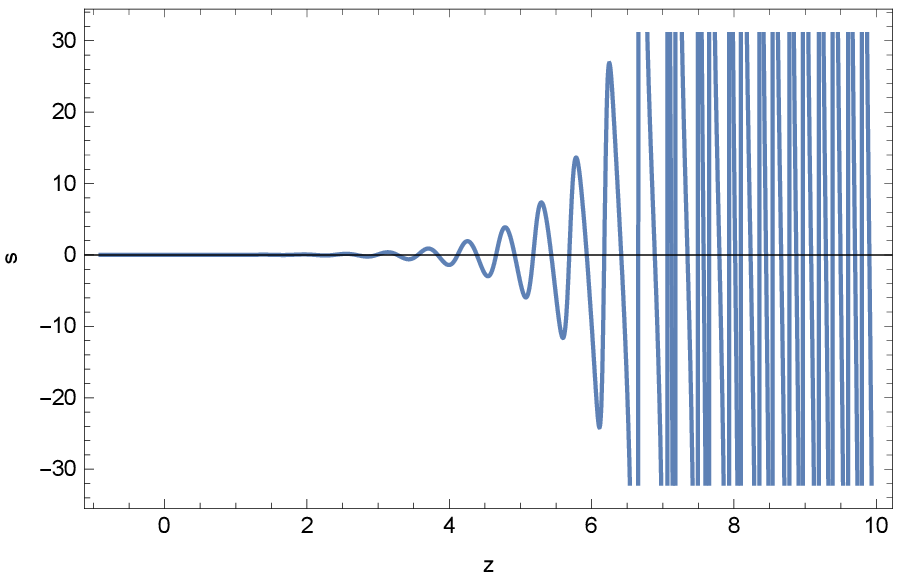}
\includegraphics[width=20pc]{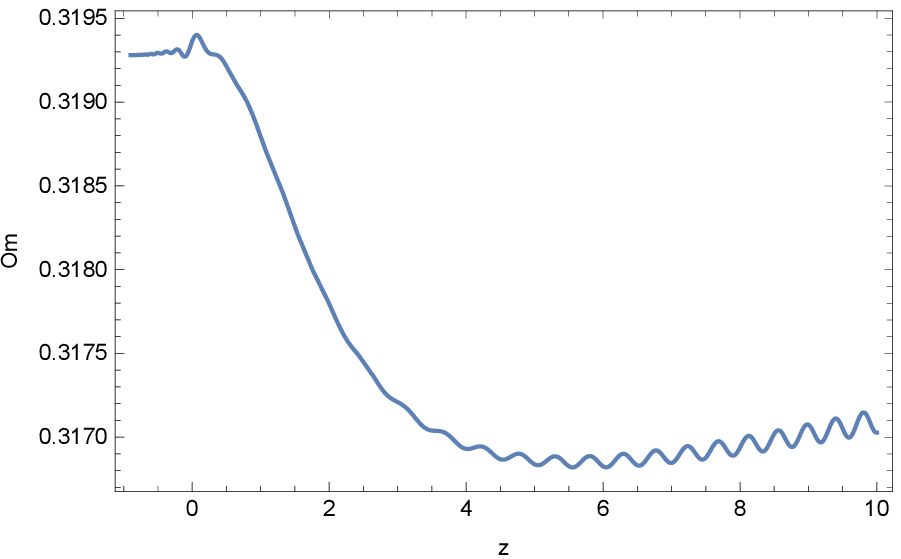}
\caption{Cosmological parameters $q$ (upper left), $j$ (upper right), $s$, (bottom left) and $s$ (bottom right) as functions of redshift. Once again, the same results as in the pure $f(R)$ case are obtained.}
\end{figure}
where $\gamma$ a dimensionless parameter, $\Lambda$ a constant
with mass dimensions $[m]^2$, $M=1.5\cdot10^{-5}\frac{50}{N}M_P$
with $N$ being the e-folding number and $\delta$ an exponent which
satisfies the relation $0<\delta<1$, see Refs.
\cite{Odintsov:2020qyw,Odintsov:2020nwm} for details. In this
particular case we assumed for simplicity that the scalar
potential is absent, as we already mentioned. In this general
framework however, there is no physical constraint that connects
the scalar potential and the scalar coupling function,
nevertheless if one takes into account the primordial
gravitational wave speed constraints, these two scalar functions
are interconnected fundamentally. For the moment though we assume
that these can be freely chosen given that no constraints on the
speed of gravitational waves are imposed, i.e $c_T^2$ does not
necessarily coincide with unity.  The $f(R)$ model we chose, was
chosen simply because it is capable of uniting early with late
time acceleration era of our Universe, due to the fact that for
large $R$, $R^2$ becomes dominant whereas for $R\to0$, $R^\delta$
becomes the dominant term, see Ref.
\cite{Odintsov:2020qyw,Odintsov:2020nwm} for a detailed analysis
on this issue. It is therefore interesting to examine whether the
addition of a scalar field can alter the dynamics of such model.
Essentially, such a model predicts non negligible dark energy
oscillations for $z\ge5$ for the statefinder parameter $y_H(z)$,
which become even more dominant in higher order derivatives.
Recently, it was showcased that the addition of a function
depending on the Gauss-Bonnet topological invariant $\mathcal{G}$,
which alone describes an oscillation-free late-time era, cannot
nullify such oscillations on the $f(R)$ model, implying that the
later is more dominant \cite{newpaper}. It is therefore sensible
to try and examine whether the addition of a canonical scalar
field coupled to the Gauss-Bonnet topological invariant can
achieve such phenomenological behavior. Essentially, by using the
same parameters for the $f(R)$ gravity as in Ref.
\cite{Odintsov:2020nwm}, meaning that $\gamma=2$,
$\Lambda=1,1895\cdot10^{-66}$eV$^2$, $\delta=\frac{1}{100}$,
$N=60$ with the initial conditions chosen as
$y_H(z=10)=\frac{\Lambda}{3m_s^2}\left(1+\frac{1+z_f}{1000}\right)$,
$\frac{dy_H}{dz}\Big|_{z=10}=\frac{\Lambda}{3m_s^2}\frac{1}{1000}$,
$\phi(z=10)=10^{-16}M_p$,
$\frac{d\phi}{dz}\Big|_{z=10}=-10^{-17}M_p$, then by solving
numerically equations (\ref{motion1} ) and (\ref{motion3}) in the
interval $[z_i,z_f]=[-0.9,10]$ with respect to $y_H$, $\phi$, it
becomes apparent that simply adding a canonical scalar field
cannot negate the dark energy oscillations. This result seems to
be in agreement with the one obtained in Ref. \cite{newpaper} for
the $f(R)+g(\mathcal{G})$ case given that $\mathcal{G}$ is quite
small in terms of the rest of the parameters. The results of our
numerical analysis for the particular model at hand can be found
in Figs. 1, 2 and 3, while in Table I we compare the values of
several statefinder quantities at present time with the
corresponding values of the $\Lambda$CDM model and we confront the
values of the dark energy density parameter $\Omega_{DE}(0)$ and
the dark energy EoS parameter $\omega_{DE}(0)$ with the latest
constraints of the Planck 2018 collaboration on cosmological
parameters \cite{Aghanim:2018eyx}. As it can be seen from Table I,
the resulting cosmological quantities and the statefinder values
at present time corresponding to the model at hand are quite close
to the $\Lambda$CDM values, and both $\Omega_{DE}(0)$ and
$\omega_{DE}(0)$ are compatible with the observational data. In
Fig. 1 we present the behavior of the statefinder $y_H$ (left) and
$\phi$ (right) as functions of the redshift, for the model at
hand. The main difference with the pure $f(R)$ gravity seems to be
the scalar field which does not oscillate. In general, the
addition of a canonical scalar field and a linear Gauss-Bonnet
topological invariant coupled to a scalar function do not suffice
to nullify dark energy oscillation at large redshifts. Also in
Fig. 2 we present the behavior of the cosmological statefinder
quantities $q$ (upper left), $j$ (upper right), $s$, (bottom left)
and $s$ (bottom right) as functions of redshift. Once again, the
same qualitative behavior as in the pure $f(R)$ case are obtained.
Finally, in Fig. 3 we present the dark energy variables, namely
the EoS (left) and the dark energy density parameter $\Omega_{DE}$
(right) as functions of the redshift. Out of these two parameters,
only the latter is free of oscillations, however neither the
canonical scalar field nor the Gauss-Bonnet topological invariant
are responsible for such feature. The $f(R)$ contribution is the
dominant term as it can also be inferred from the rest results.
\begin{table}[h!]
\label{Table1}
\caption{}
\begin{center}
\begin{tabular}{|r|r|r|}
\hline
\textbf{Parameter}&\textbf{$f(R)$}&\textbf{$\Lambda$CDM Value}\\ \hline
q(z=0)&-0.520954&-0.535\\ \hline
j(z=0)&1.00319&1\\ \hline
s(z=0)&-0.00104169&0\\ \hline
$Om(z=0)$&0.319364&0.3153$\pm$0.07\\ \hline
$\Omega_{DE}(0)$&0.683948&0.6847$\pm$0.0073\\ \hline
$\omega_{DE}(0)$&-0.995205&-1.018$\pm$0.031\\ \hline
\end{tabular}
\end{center}
\end{table}

\begin{figure}[h!]
\centering
\label{plot3}
\includegraphics[width=20pc]{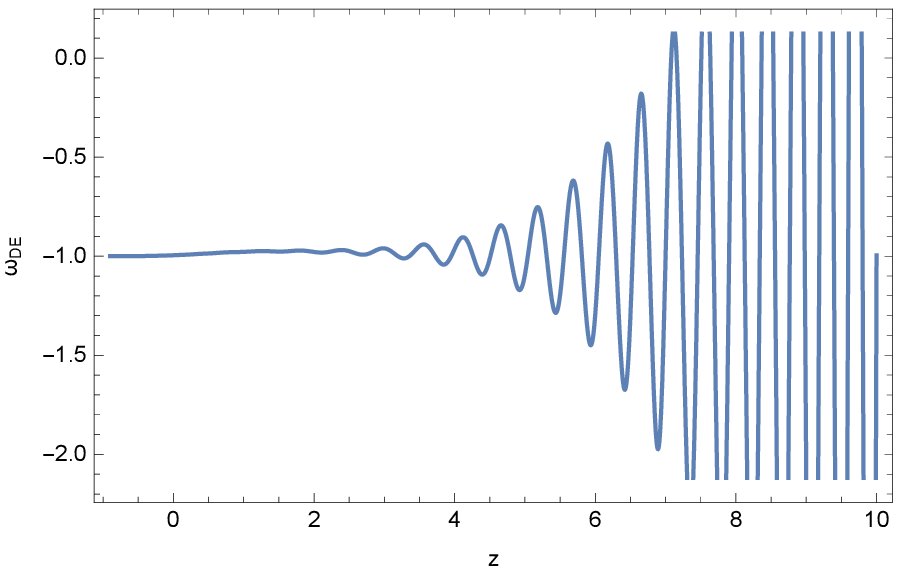}
\includegraphics[width=20pc]{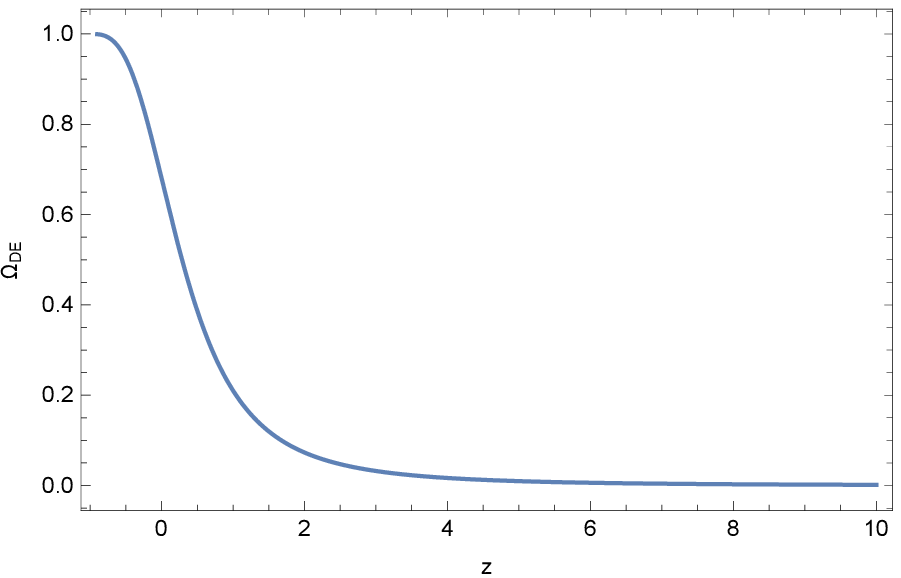}
\caption{Dark energy variables, the EoS (left) and the Density
parameter $\Omega_{DE}$ (right). Out of these two parameters, only
the latter is free of oscillations however neither the canonical
scalar field nor the Gauss-Bonnet topological invariant are
responsible for such feature. The $f(R)$ contribution is the
dominant term as it can also be inferred from the rest results.}
\end{figure}
Thus for this particular class of potential-less models, the
$f(R)$ gravity part seems to dominate the late-time evolution. In
the following sections we shall also introduce a potential, and in
parallel we shall assume an Einstein-Hilbert $f(R)$ term. At a
later section we shall constrain the functional forms of the
scalar potential and the scalar coupling function, in order to see
how the late-time dynamics are affected by these changes.

As a final comment, we should mention that even though the value
of the scalar field seems to increase with respect to time, the
rate of increase is smaller and subdominant when it is compared to
the rate of the $f(R)$ gravity terms, and in particular from the
term $\sim R^\delta$. Subsequently, the $f(R)$ part is dominant in
comparison to the scalar terms and thus the increasing value of
$\phi$ does not contradict the overall phenomenology. For this
exact reason, one observes dark energy oscillations in the high
redshift area, a feature which arises in the pure $f(R)$ case.

\section{Einstein-Gauss-Bonnet Gravity In the Presence of a Scalar Potential}

Let us now proceed with a different approach. We shall assume that
the $f(R)$ case is reduced to a simple $f(R)=R$ case and that the
scalar potential is now present in the formalism. Let us assume
that the potential has the following arbitrary form,
\begin{equation}
\centering
\label{VB}
V(\phi)=\left(\frac{\phi}{M_P}\right)^4\, ,
\end{equation}
\begin{figure}[h!]
\centering
\label{plot4}
\includegraphics[width=20pc]{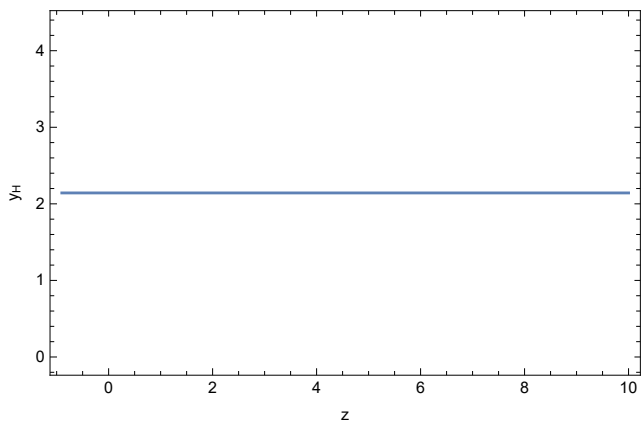}
\caption{Statefinder $y_H$ for the $R$ case in the presence of a
scalar potential.}
\end{figure}
while the scalar coupling function has the following form,
\begin{equation}
\centering
\label{xiB}
\xi(\phi)=\left(\frac{\phi}{M_P}\right)^2\, ,
\end{equation}
\begin{figure}[h!]
\centering
\label{plot5}
\includegraphics[width=20pc]{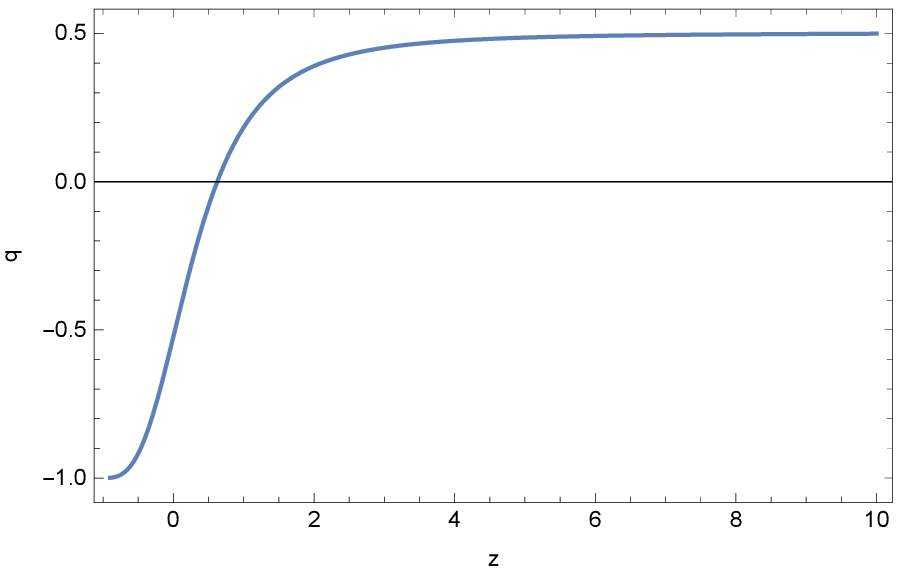}
\includegraphics[width=20pc]{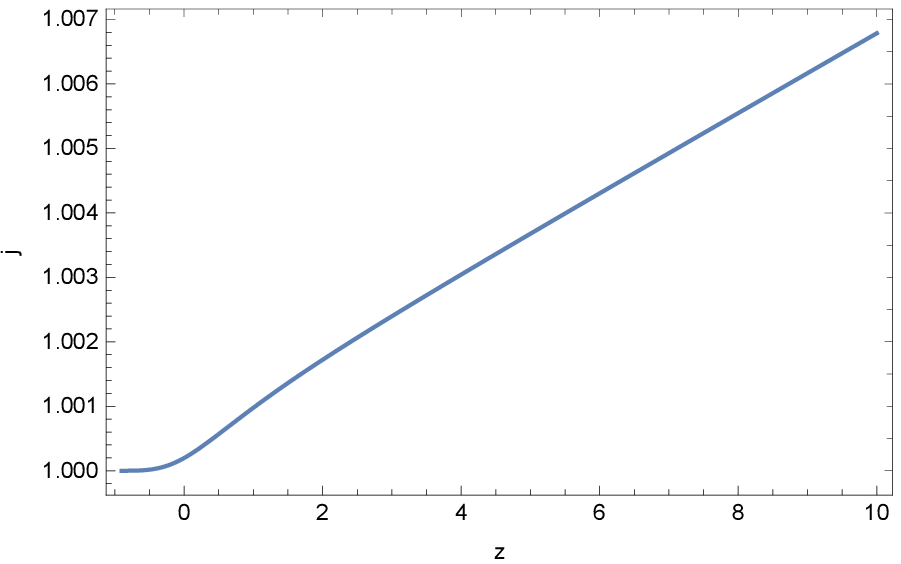}
\includegraphics[width=20pc]{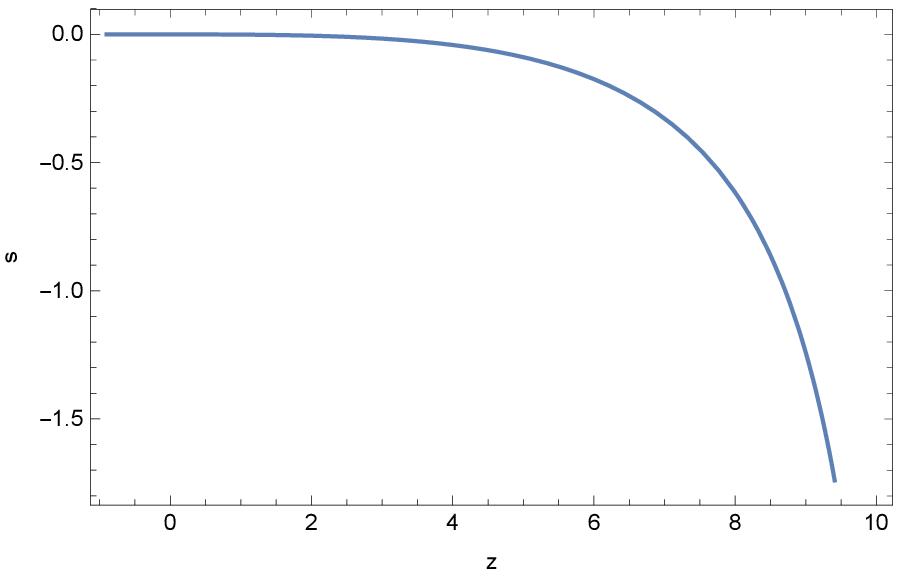}
\includegraphics[width=20pc]{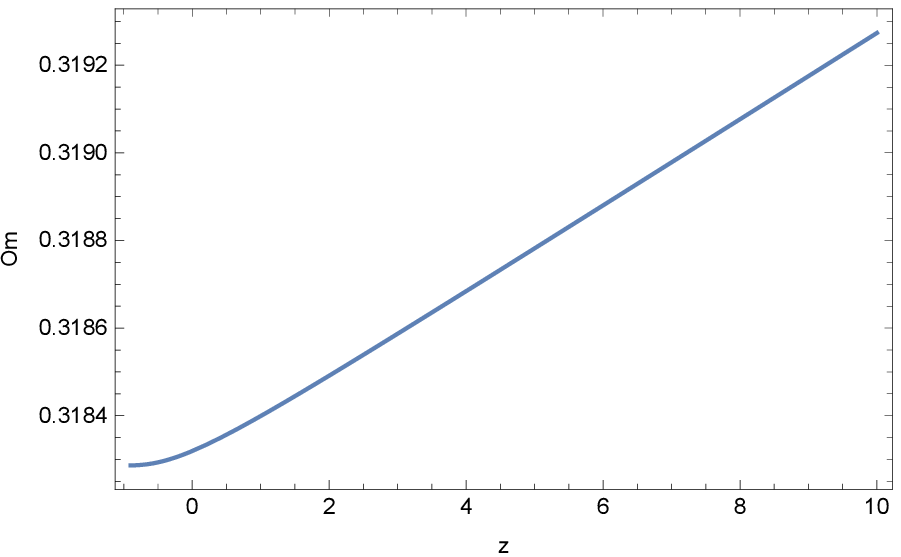}
\caption{ Deceleration $q$ (upper left), jerk $j$ (upper right),
snap $s$ (bottom left) and $Om$ (bottom right) as functions of
redshift. In this case, no dark energy oscillations are present
since such feature is generated from additional $R$ terms. All
variables are in agreement with the $\Lambda$CDM however what is
fascinating is the snap parameter which decreases with time.}
\end{figure}
In this case we shall assume simple power-law models for the
scalar functions which are normalized with respect to the reduced
Planck mass. Such models are frequently used in the inflationary
era where the slow-roll conditions for the scalar field are
usually assumed to hold true. In the late time era, there exists
no need to apply the slow-roll conditions since they do not hold
true. Let us proceed with the numerical results. In this case, the
Einstein-Hilbert form of $f(R)$ implies that certain terms in
Eq.(\ref{motion1}) are discarded which facilitates our study.
Furthermore, since $\dot R$ is now absent, there exists no second
derivative of statefinder $y_H$. In fact, if it was not for the
scalar field which has a term $\ddot\phi$ proportional to $y_H'$
in the continuity equation, the aforementioned statefinder
function would need no initial conditions to be specified. Here,
we shall only assume that $y_H(z=10)$ is once again equal to
$y_H(z=10)=\frac{\Lambda}{3m_s^2}\left(1+\frac{1+z_f}{1000}\right)$
and in addition, $\phi(z=10)=M_P$,
$\frac{d\phi}{dz}\Big|_{z=10}=M_P$ then the results of our
analysis can be found in Figs. 4, 5 and
6. Also in Table II, as in the model of the
previous section, we compare the values of several statefinder
quantities at present time with the corresponding values of the
$\Lambda$CDM model and we confront the values of the dark energy
density parameter $\Omega_{DE}(0)$ and the dark energy EoS
parameter $\omega_{DE}(0)$ with the latest constraints of the
Planck 2018 collaboration on cosmological parameters
\cite{Aghanim:2018eyx}. As it can be seen from Table II,
the resulting cosmological quantities and the statefinder values
at present time corresponding to the model at hand are quite close
to the $\Lambda$CDM values, and both $\Omega_{DE}(0)$ and
$\omega_{DE}(0)$ are compatible with the observational data.
\begin{table}[h!]
\label{Table2}
\caption{}
\begin{center}
\begin{tabular}{|r|r|r|}
\hline
\textbf{Parameter}&\textbf{$R$}&\textbf{$\Lambda$CDM Value}\\ \hline
q(z=0)&-0.522521&-0.535\\ \hline
j(z=0)&1.0002&1\\ \hline
s(z=0)&-0.00006431&0\\ \hline
$Om(z=0)$&0.318319&0.3153$\pm$0.07\\ \hline
$\Omega_{DE}(0)$&0.681713&0.6847$\pm$0.0073\\ \hline
$\omega_{DE}(0)$&-1&-1.018$\pm$0.031\\ \hline
\end{tabular}
\end{center}
\end{table}
It can easily be inferred from the plots that the qualitative
behavior of the model under study is quite close to the
$\Lambda$CDM model. One striking feature is that the dark energy
density $\rho_{DE}$ is nearly constant throughout the interval
[-0.9,10], as indicated by $y_H$, and therefore the EoS parameter
$\omega_{DE}$ is also nearly equal to $-1$. This result is robust
towards changing the free parameters for this particular model.
The rest of the cosmological parameters however seem to have an
infinitesimal evolution, for instance the jerk parameter is quite
close to $j=1$ but not exactly equal to unity as $\omega_{DE}$ is
$-1$.
\begin{figure}[h!]
\centering
\label{plot6}
\includegraphics[width=20pc]{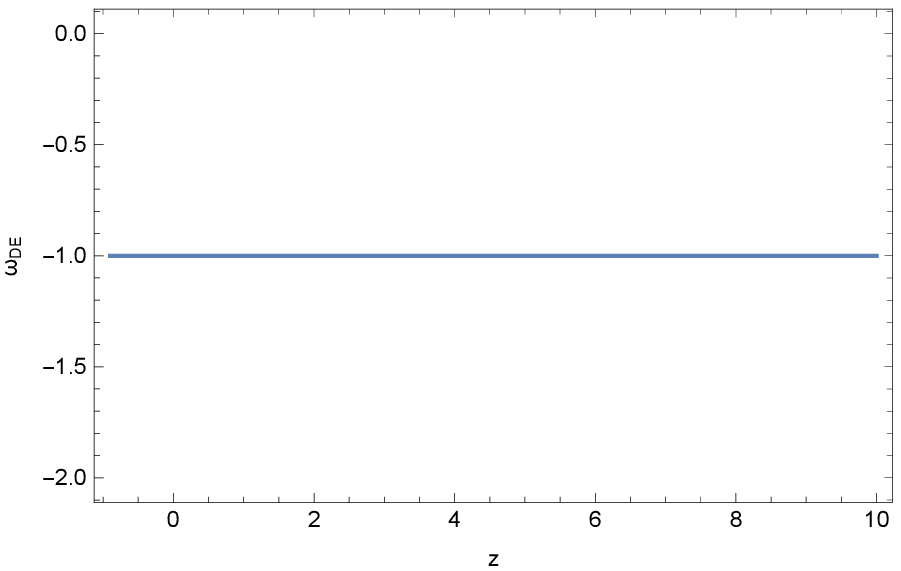}
\includegraphics[width=20pc]{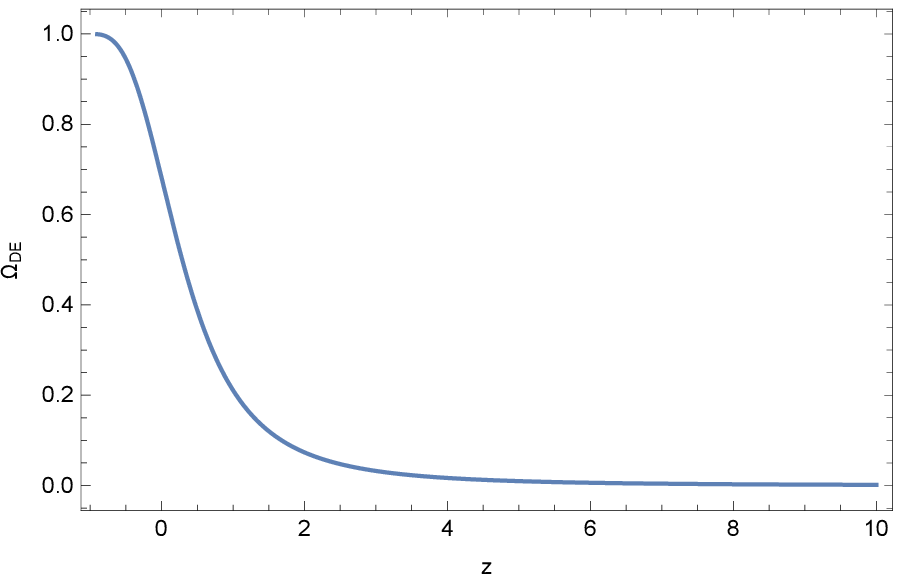}
\caption{Equation of state parameter $\omega_{DE}$ and density
parameter $\Omega_{DE}$. Due to the fact that statefinder $y_H$ is
numerically found to be slowly evolving or nearly constant during
the stage of our Universe's evolution, the EoS is subsequently
exactly nearly equal to $\omega_{DE}=-1$.}
\end{figure}
As a final comment, it should be noted that both models studied so
far do not have a fixed value for the velocity of the primordial
gravitational waves these models produce. Since
\cite{Hwang:2005hb,Oikonomou:2020oil,Odintsov:2020xji,Oikonomou:2020sij,Odintsov:2020sqy,Odintsov:2020zkl},
\begin{equation}
\centering
\label{cT}
c_T^2=1-\frac{Q_f}{2Q_t}\, ,
\end{equation}
\begin{figure}[h!]
\centering
\label{plot7}
\includegraphics[width=20pc]{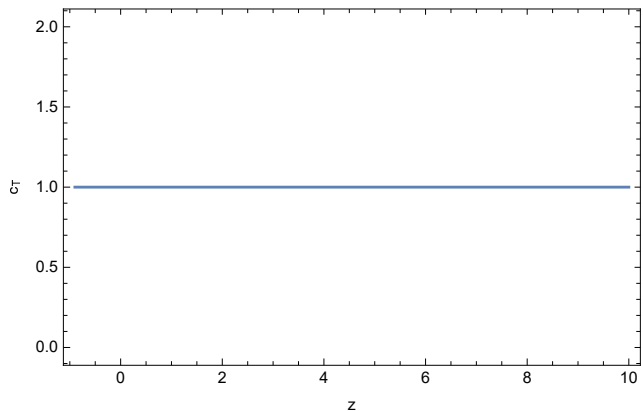}
\caption{Gravitational wave velocity as function of redshift. The
main result seems to be a constant value which in turn implies
that since $Q_f\ll1$, the scalar coupling function $\xi(\phi)$
satisfies the differential equation $\ddot\xi=H\dot\xi$.}
\end{figure}
with $Q_f=16(\ddot\xi-H\dot\xi)$ and $Q_t=M_P^2-8\dot\xi H$, then
it stands to reason that the velocity obtains arbitrary values.
Despite being arbitrary, its value is essentially equal to unity
due to the fact that $Q_f\ll Q_t$, and thus an infinitesimal value
is subtracted from unity in Eq. (\ref{cT}). In particular,
$Q_f\sim\mathcal{O}(10^{-33})$ and $Q_t\sim\mathcal{O}(10^{36})$
hence the reason why the velocity is equal to unity, and this can
also be seen in Fig. 7, but the fact that $Q_f\ll1$,
even in Planck units where $\kappa=1$, implies that a relation of
the form $Q_f=0$ or $\ddot\xi=H\dot\xi$ for the Gauss-Bonnet
scalar coupling function is not a random choice. In the next
section we shall examine the phenomenological implications of
constraining the aforementioned velocity be letting $Q_f=0$.

\section{Phenomenology with the Constraint $\ddot\xi=H\dot\xi$}

The above formulation seems to be in general in good agreement
with not only the observational data, but also with the
$\Lambda$CDM model itself. As stressed in the last model however,
the Einstein-Gauss-Bonnet models have a flaw, having to do with a
production of a primordial tensor power spectrum, with propagation
speed different from unity. The primordial gravitational wave
speed however, must be equal to that of light's  in order to
comply with the GW170817 event \cite{GBM:2017lvd}. The deviation
from unity for the above models is perhaps small in magnitude,
implying that the effective value is $c_T\simeq1$ however is it
interesting to examine the cosmological implications on the
late-time era when constraints on the velocity of tensor
perturbations are imposed. Subsequently, following Ref.
\cite{Oikonomou:2020oil,Odintsov:2020xji,Oikonomou:2020sij,Odintsov:2020sqy,Odintsov:2020zkl}
we shall proceed by assuming $\ddot\xi=H\dot\xi$. During the
inflationary era, where the slow-roll conditions are assumed to
hold true, the previous differential equation can define the time
evolution of the scalar field, $\dot\phi$. This study though was
performed using the slow-roll assumptions, which of course do not
hold true in the present late-time context. Although one can
easily imply that the scalar functions of the model continue to
have the same primordial relation they had during the inflationary
era, and also that the gravitational wave speed remains unity
after the horizon crossing of the primordial tensor modes, here we
shall adopt a different approach, and since the redshift is used
as a variable, we shall solve analytically the equation. Since
$\ddot\xi=H\dot\xi$, the solution reads,
\begin{equation}
\centering
\label{dotxi}
\dot\xi=\lambda e^{\int{Hdt}}\, ,
\end{equation}
where $\lambda$ is an integration constant. Since the definition
of redshift is $\frac{dz}{dt}=-H(1+z)$, the above integral can be
solved analytically, and thus the final solution is written as,
\begin{equation}
\centering \label{dotxi2}
\dot\xi=a(t)\lambda=\frac{\lambda}{1+z}\, .
\end{equation}
This is a quite useful result, to say the least, since by simply
imposing constraints on the velocity of gravitational waves in the
late-time era, the degrees of freedom of the model are decreased
by one, similar to the inflationary era, and obtain a functional
constraint on $\dot\xi$ which seems to be model independent. In
contrast to the previous sections, no definition for $\xi(\phi)$
is needed since essentially a transformation was performed which
replaced $\dot\xi(\phi)$ with $\dot\xi(z)$, and given that in the
equations (\ref{motion1}) and (\ref{motion3}) which we aim to
solve numerically, only $\dot\xi$ is present, the overall
phenomenology is now significantly altered. Now, the equations of
motion are altered as shown below,
\begin{equation}
\centering
\label{motion8}
\frac{3f_RH^2}{\kappa^2}=\rho_m+\frac{1}{2}\dot\phi^2+V+\frac{f_R R-f}{2\kappa^2}-\frac{3H\dot f_R}{\kappa^2}+24\frac{\lambda}{1+z}H^3\, ,
\end{equation}
\begin{equation}
\centering
\label{motion9}
-\frac{2f_R\dot H}{\kappa^2}=\rho_m+P_m+\dot\phi^2+\frac{\ddot f_R-H\dot f_R}{\kappa^2}-16\frac{\lambda}{1+z}H\dot H\, ,
\end{equation}
\begin{equation}
\centering \label{motion10}
V_\phi+\ddot\phi+3H\dot\phi-\frac{f_\phi}{2\kappa^2}+\frac{\lambda}{1+z}\frac{\mathcal{G}}{\dot\phi}=0\,
.
\end{equation}
It should be noted that all the previous equations acquired in
section II are still valid even when the constraint is applied.
Furthermore, given that $\mathcal{G}$ is small from its nature,
for certain values of $\lambda$, the phenomenology for such choice
is indistinguishable from the one obtained without the constraint.
For instance, if we recall the results for the $f(R)$ model
studied previously, it was mentioned that the scalar field cannot
alter the results. The same can be said about the case of $f(R)$
with $\dot\xi=\frac{\lambda}{1+z}$ for a plethora of values for
$\lambda$. By altering $\lambda$ and giving it a quite large
value, say $\lambda=10^{100}$, then the solution diverges and
compatibility cannot be achieved. Let us proceed with a specific
model and examine the impact the constraint has on the late-time
evolution.

Consider a non-minimally coupled model of the form,
\begin{equation}
\centering
\label{hR}
f(R,\phi)=h(\phi)R\, ,
\end{equation}
\begin{equation}
\centering
\label{h}
h(\phi)=\frac{\phi_0}{\phi}\, ,
\end{equation}
\begin{equation}
\centering
\label{VC}
V(\phi)=\frac{V_0}{\phi}\, ,
\end{equation}
and obviously
\begin{equation}
\centering
\label{xiC}
\dot\xi=\frac{\lambda}{1+z}\, ,
\end{equation}

\begin{figure}[h!]
\centering
\label{plot9}
\includegraphics[width=20pc]{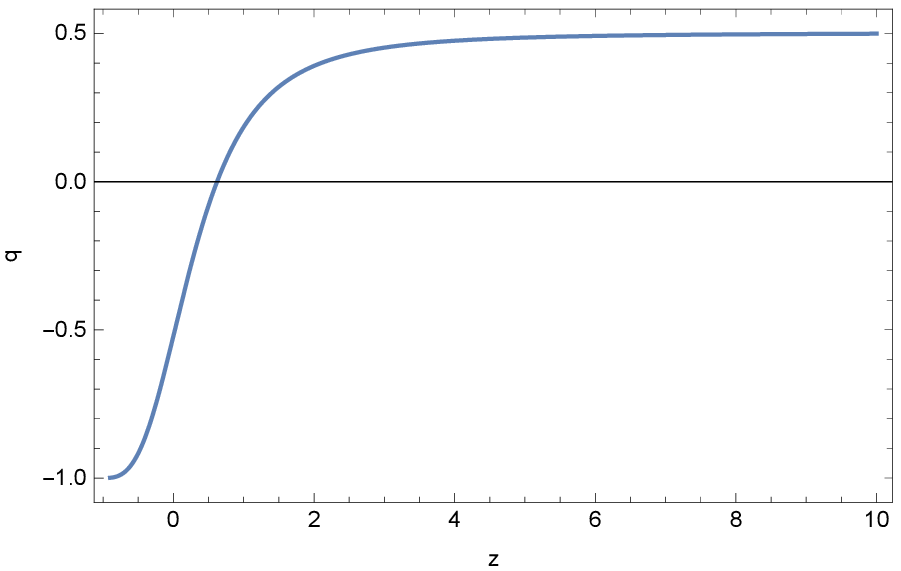}
\includegraphics[width=20pc]{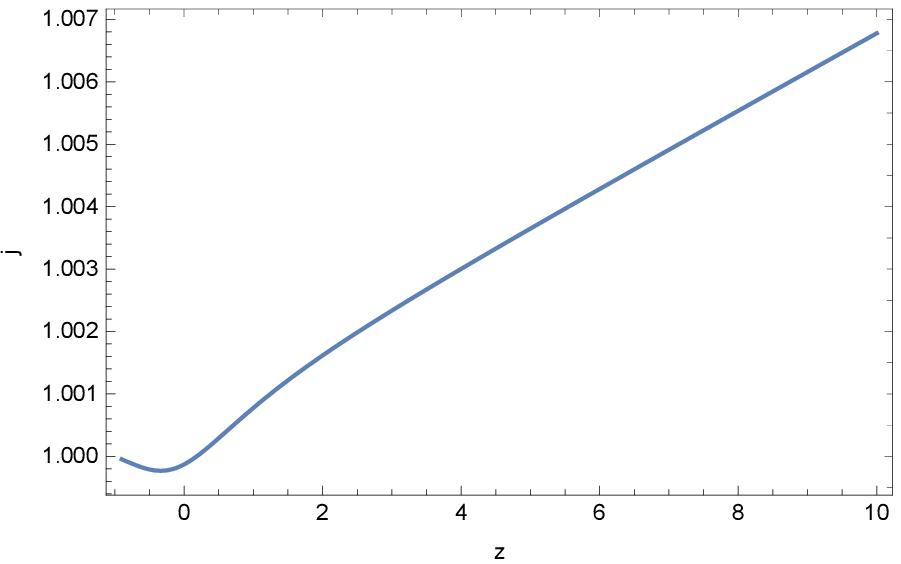}
\includegraphics[width=20pc]{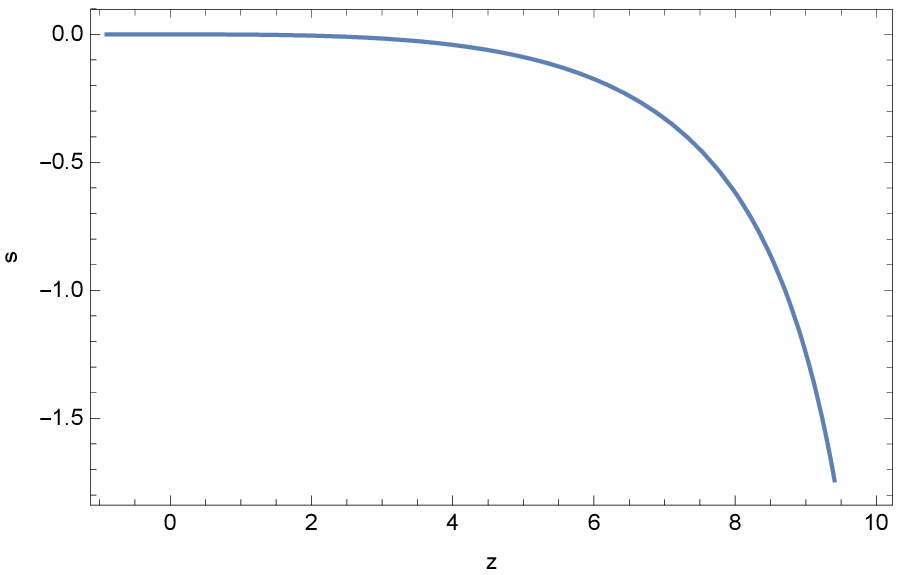}
\includegraphics[width=20pc]{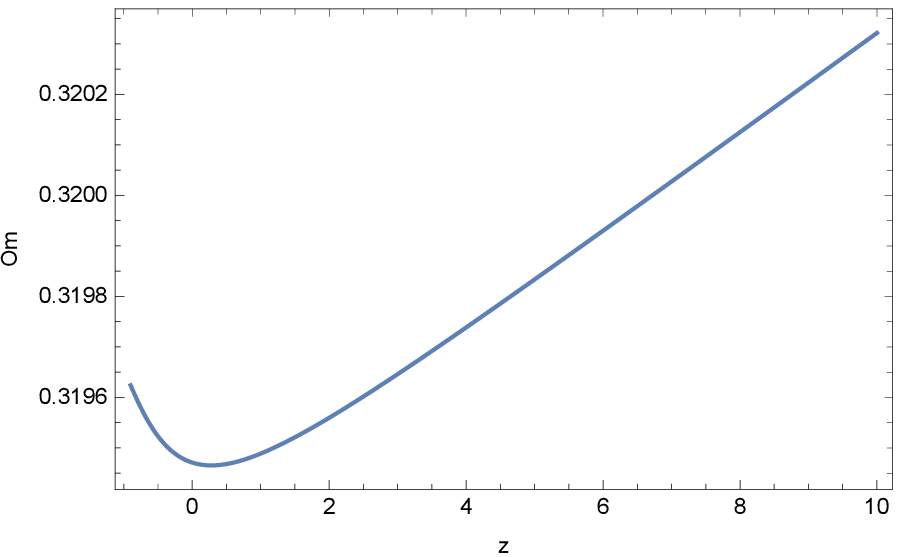}
\caption{Cosmological parameters $q$( upper left), $j$ (upper right), $s$ (bottom left) and $Om$ (bottom right) with respect to $z$. Once again, no dramatic change is present in the non minimally coupled constrained case with the $R$ unconstrained.}
\end{figure}
where $\phi_0$ and $V_0$ are auxiliary parameters with mass
dimensions $[m]$ and $[m]^5$ respectively. In this case as well,
since there exists only a linear $R$ term, only a single initial
condition for $y_H$ is needed. As was the case with the previous
two models, we shall use the same value, meaning
$y_H(z=10)=\frac{\Lambda}{3m_s^2}\left(1+\frac{1+z_f}{1000}\right)$.
In consequence, letting $\phi_0=1$, $V_0=1$, $\lambda=1$,
$\phi(z=10)=10^{-35}M_P$,
$\frac{d\phi}{dz}\Big|_{z=10}=-10^{-20}M_P$ then the results
obtained are compatible with the $\Lambda$CDM model as shown in
Fig. 9, while in Fig. 10 we present the
behavior of the dark energy EoS parameter and the dark energy
density parameter as functions of the redshift. The results of our
numerical analysis corresponding to the values of the statefinders
and of the dark energy EoS parameter and the dark energy density
parameter at present time, can be found in Table III. As
it can be seen in Table III our model is in good
qualitative agreement with the $\Lambda$CDM model and also is
compatible with the 2018 Planck constraints on the cosmological
parameters, when the dark energy EoS parameter and the dark energy
density parameter are considered.
\begin{table}[h!]
\label{Table3}
\caption{}
\begin{center}
\begin{tabular}{|r|r|r|}
\hline
\textbf{Parameter}&\textbf{$h(\phi)R$}&\textbf{$\Lambda$CDM Value}\\ \hline
q(z=0)&-0.520794&-0.535\\ \hline
j(z=0)&0.99987&1\\ \hline
s(z=0)&-0.00004423&0\\ \hline
$Om(z=0)$&0.319364&0.3153$\pm$0.07\\ \hline
$\Omega_{DE}(0)$&0.680671&0.6847$\pm$0.0073\\ \hline
$\omega_{DE}(0)$&-0.99984&-1.018$\pm$0.031\\ \hline
\end{tabular}
\end{center}
\end{table}

\begin{figure}[h!]
\centering
\label{plot10}
\includegraphics[width=20pc]{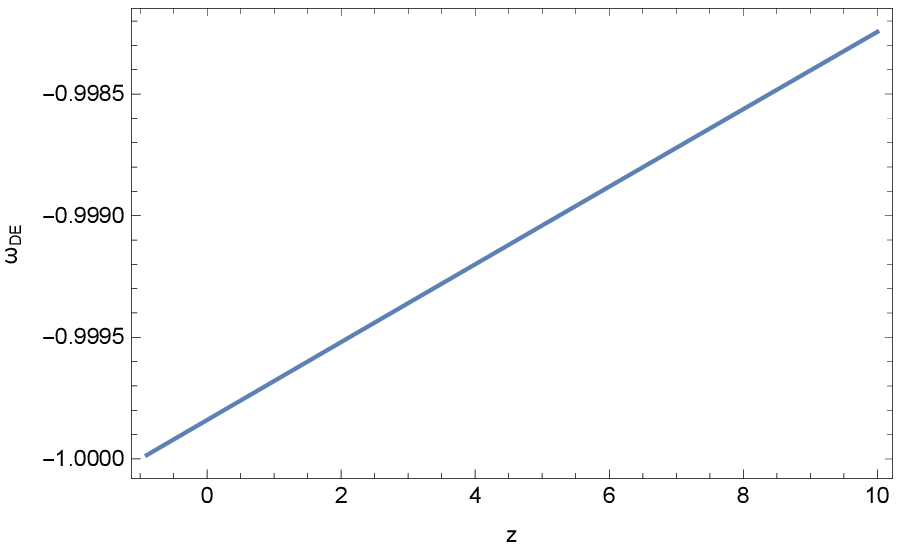}
\includegraphics[width=20pc]{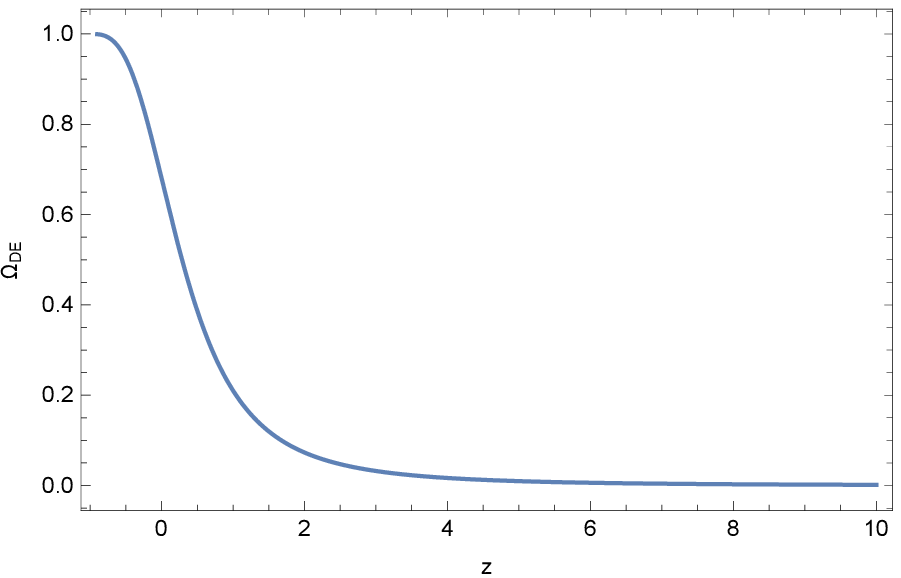}
\caption{Dark energy parameters $\omega_{DE}$ (left) and
$\Omega_{DE}$ for the non minimally coupled model. Since $y_H$ is
depending on $z$, $\omega_{DE}$ evolves as well however
infinitesimally near the expected value $-1$ while $\Omega_{DE}$
increases with time as it is expected.}
\end{figure}
Genuinely speaking, the linear $R$ case we studied in the previous
section, without constraints and the $h(\phi)R$ case with
constraints studied in this section, do not differ so much. The
main difference lies in the evolution of statefinder $y_H$ and
subsequently the EoS on the latter case, however even when the EoS
is dynamically evolving, it does so with an infinitesimal rate
that its value essentially cannot be distinguished from unity.

As a comment, it should be noted that in this case no difference
between the constrained and the unconstrained Gauss-Bonnet
phenomenology is found, however the latter seems quite arbitrary
from one perspective, meaning that the velocity of gravitational
waves is dynamically evolving in various cosmological eras and
$c_T$ just happens to be unity since $Q_f\ll Q_t$. However, even
uncontrolled, the fact that $Q_f\ll1$ even in Planck units implies
that the relation $\ddot\xi=H\dot\xi$ is satisfied one way or
another, hence instead of coming to such numerical conclusion at
the end, it is beneficial to begin with such statement as the
degrees of freedom seem to decrease in the first place. The main
idea was to examine a model with the constraint $c_T^2=1$ and
prove that compatibility can be achieved by taking into
consideration that the constraint imposed from the velocity of
gravitational waves decreases the degrees of freedom such that the
Gauss-Bonnet scalar coupling function can be replaced by a single
parameter. In the literature \cite{Kleidis:2019ywv} such a
question has been addressed, and the results were quite different
quantitatively in comparison to the present model. In principle
however, the compatibility is a model dependent feature.

Before closing we need to discuss an interesting scenario, in view
of the unified description of inflation with the dark energy era
that Einstein-Gauss-Bonnet theory, combined with the fact that it
is a string theory originating theory. In the present paper we
assumed that the dark matter perfect fluid consists of an unknown
particle, but string theory has also offered the possibility of
having axion like particles present even in the pre-inflationary
era. In fact, in the axion like particle phenomenology with a
primordial pre-inflationary era broken $U(1)$ symmetry. The axion
due to the breaking of this symmetry is frozen in its vacuum
expectation value, but as the Universe expands, the axion behaves
as a condensate and evolves as a dark matter perfect fluid, which
makes it a perfect candidate for a low-mass weakly interactive
massive dark matter particle. Such scenarios in the context of
modified gravity have been studied in the literature
\cite{Odintsov:2020iui,Nojiri:2020pqr,Odintsov:2020nwm,Nojiri:2019riz},
so one interesting scenario is to have the combined presence of
the axion coupled to the Gauss-Bonnet scalar. This would utterly
change the symmetry breaking patter of the primordial $U(1)$
symmetry, due to the presence of the coupling $\xi(\phi)$ in the
axion equation of motion, even pre-inflationary. The calculation
might get easier if it is assumed that the Gauss-Bonnet
corrections and the flat four dimensional spacetime, FRW-like, are
the resulting outcomes of the quantum era. Also, the presence of
the Gauss-Bonnet non-minimal coupling would alter the
post-inflationary evolution, and in addition, in this scenario,
the axion $U(1)$ symmetry might be unbroken during the
inflationary era. These issues are interesting material for a
focused future work.

\section{Conclusions}

In this work we investigated the late-time phenomenology aspects
of scalar-coupled  $f(R,\mathcal{G})$ gravity. We focused on
theories of Einstein-Gauss-Bonnet form, and we examined three
types of models, $f(R)$ gravity Einstein-Gauss-Bonnet models, and
pure Einstein-Gauss-Bonnet models, with arbitrary choice of the
scalar functions of the models and with constrained functions of
the models. Our numerical analysis indicated that for the models
containing the $f(R)$ gravity, the late-time dynamics is very much
affected by the $f(R)$ gravity part, and thus in those cases, the
Einstein-Gauss-Bonnet coupling does not affect the dynamics. For
the pure Einstein-Gauss-Bonnet, we made a novel assumption related
to the  requirement that the primordial gravitational wave speed
is equal to unity, which in turn imposed a functional constraint
on the functional form of the Gauss-Bonnet scalar coupling
function. The exiting feature in the late-time study by taking
into account the gravitational wave speed constraint, is the fact
that the functional form of $\dot{\xi}$ is model independent, and
has a specific form given in terms of the Hubble rate and the
redshift. This simplification is rather interesting to think that
there is a strong motivation to assume that the primordial gravity
wave speed should be set equal to unity for all the cosmic times
after the first horizon crossing of the primordial tensor modes.
This could in fact constrain the functional forms of the scalar
potential and of the scalar coupling function, directly from the
inflationary era and thereafter, but we did not go to deep to this
study, since it seems that the Gauss-Bonnet coupling as we studied
it in section II does not play a significant role during the
late-time era. It should actually, it is of the order $\sim H^4$,
but we aimed to formally investigate the phenomenology of these
models. The positive outcome we keep is that the primordial
gravitational wave speed constraint has some effect on the
late-time cosmological dynamics, and this is a motivation for us
to go investigate astrophysical scenarios related to
Einstein-Gauss-Bonnet models, with the potential and the scalar
coupling functions being related in the way we demonstrated in
\cite{Oikonomou:2020oil,Odintsov:2020xji,Oikonomou:2020sij,Odintsov:2020sqy,Odintsov:2020zkl}.
Such task we aim to address in a future work.

Let us discuss at this point the issue of choosing the scalar
functions of the models we studied in this paper. In order to
address this comment we need to elaborate further on the concept
of gravitational waves. Theories containing a Gauss-Bonnet term
coupled to an arbitrary scalar function are notorious for
producing primordial gravitational waves which propagate with a
velocity which is different from the speed of light. Our initial
work on Gauss-Bonnet started by confronting this nasty feature
during the inflationary era and coming up with ways in order to
remedy the theory, see Ref. \cite{Odintsov:2020sqy}. The reason
that the theory needs to be rectified, from our point of view, it
that there exists no known mechanism which could produce massive
primordial gravitons which were later turned into massless
according to the recent GW170817 event. Therefore, a reasonable
assumption is to impose that the theory is described by massless
gravitons throughout the evolution of the Universe. The main
result extracted from this simple statement is that the
Gauss-Bonnet scalar coupling function must satisfy the
differential equation $\ddot\xi=H\dot\xi$ and since Hubble's
parameter is given from the Friedmann equation, it becomes
apparent that in this scenario both scalar coupling functions are
connected, i.e. $V(\phi)$ and $\xi(\phi)$ are chosen in such a way
so that the aforementioned differential equation is satisfied
properly. This can be seen easily in the slow-roll regime where
many terms are assumed to be subleading but during the late-time
era, we are interested in all of them. The main idea was to start
without imposing the constraint derived from the velocity of
gravitational waves and afterwards compare the results between a
constrained and an unconstrained model. As it turns out, both
assumptions lead to relative similar results given that in the set
of equations, other terms are dominant during the late-time era.
Moreover, in order to perform a fully self-consistent study,
further auxiliary parameters such as the jerk, snap and $Om(z)$
where used in order to see the pros and cons of the models studied
in this case and to see explicitly where and if some cosmological
parameters deviate from observations, thus rendering the model
unsuitable. Concerning the initial conditions at redshift $z=10$,
the first variable is designated in order to coincide with
observations while the initial value of the scalar field, since it
cannot be observed, is chosen arbitrarily. The previous value
seems to produce a smooth evolution for the scalar field with
respect to cosmic time and moreover all the physical observables
values of the model  are compatible with the observational data.

Moreover, let us comment that in the present text, the early-time
phenomenology was not considered, however in order have an
inferior Gauss-Bonnet term in the late-time era, one would expect
that the scalar field related terms are subleading from the first
horizon crossing and thereafter. No comment can be made about the
Planck era unfortunately, but at least in the inflationary era,
the $R$ part of the equations of motion should at best receive
mild corrections so that to obtain acceptable values for the
scalar spectral index and the tensor-to-scalar ratio. Assuming
that the scalar-field evolves with either a slow-roll or a
constant rate of roll, then by implementing either similar
slow-roll conditions for the Gauss-Bonnet coupling in the
unconstrained case, or working in the constrained case and in the
presence of an arbitrary scalar potential, then compatibility with
the observational data can be achieved relatively easily, while
simultaneously the order of magnitude of several scalar functions
and their derivatives can be negligible. To summarize, a smooth
early-time description which ensures that the scalar components of
the model act as an effective dark energy density in the Friedmann
equation can be achieved by working with the classical
prescription of inflation and assume a potential driven inflation
with either slow or constant-roll evolution. In either case, the
$R$ part of the equations of motion becomes dominant in comparison
to the scalar one as time flows and especially during the
late-time era.

Before closing we need to discuss an importance issue, related
with the choice of the arbitrary functions of our models.  As we
demonstrated, the Gauss-Bonnet coupling contributes slightly to
the late-time era. The fact that the Gauss-Bonnet coupling does
not affect so significantly the late-time phenomenology of the
models studied in the unconstrained case, is a direct consequence
of their relevance. It turns out that the Gauss-Bonnet term
$\mathcal{G}$ is not dominant in the dark energy era (it scales as
$\mathcal{G}\sim H^4$ for a flat FRW spacetime). In consequence,
all the scalar terms manage to produce what is perceived as a a
nearly constant term in the Friedmann equation, which in turn is
interpreted as an effective late-time cosmological constant. In
principle, in order to see different results, one would need a
really strong contribution from the scalar components of the
model, however no such case was observed in the respective
computer program which was developed for this purpose. In fact as
we evinced, in the first case, the $f(R)$ contribution, and mainly
from the third term $R^\delta$, is the dominant driving force.
Similarly, in the second case, a simple $R$ term even in the
presence of a scalar potential is the driving force. Finally, in
the constrained model where the non-minimally coupled case was
examined, the realization that the scalar field itself is
subleading is the factor which ensures that the $R$ contribution
is once again dominant. The reason behind such designation for the
scalar coupling was mainly the simplicity of the resulting
expressions and also the production of a viable inflationary era,
which can be ascertained from recent observations, hence the
reason why power-laws and also dilatonic couplings were assumed.
Finally, it is interesting to compare the results obtained in this
paper, with the ones obtained in Ref. \cite{Paul:2008id} using
again a dilatonic Einstein-Gauss-Bonnet gravity, but in the
context of an emergent gravity scenario. The results produced in
our paper are different from the ones obtained in Ref.
\cite{Paul:2008id}. This can be attributed to the specific form of
the scale factor that the authors have chosen and/or to the non
canonical kinetic term of the scalar field. In the present
article, we did not assume a specific form for the scale factor
and in fact the scale factor is directly derivable from the
equations of motion once $V(\phi)$, $\xi(\phi)$ and $\phi(t)$ are
chosen. Moreover, the kinetic term is assumed to be canonical,
meaning that $\dot\phi$ is not coupled to $\phi$. These two
different assumptions may be the reason why the results do not
agree, since Ref. \cite{Paul:2008id} deals with the emergent
Universe scenario, which has its own attributes however, so this
issue was worthy of mentioning.

\section*{Acknowledgments}

This work was supported by MINECO (Spain), project
PID2019-104397GB-I00 and PHAROS COST Action (CA16214) (SDO).

\end{document}